\documentstyle[epsfig]{article}
\def\slantfrac#1#2{\kern1em^{#1}\kern-.3em/\kern-.1em_{#2}}
\begin{document}  
\title{IBM: parameter symmetry, hidden symmetries and 
transformations of boson operators\thanks{Talk given by A. M. Siirokov at 
XXII Symposium on Nuclear Physics, Oaxtepec, Morelos, M\'exico, 5--8
January, 1999. }}
\author{A.~M.~Shirokov$^{a}$, N.~A.~Smirnova$^{a,b}$, Yu.~F.~Smirnov$^{a,c}$, 
O.~Casta\~nos$^{c}$, A.~Frank$^{c}$\\[2mm]
$^{a}$ Institute for Nuclear Physics, Moscow State University,
119899 Moscow, Russia \\
$^{b}$GANIL, BP 5027, F-14076, Caen Cedex 5, France \\
$^{c}$Instituto de Ciencias Nucleares, UNAM, M\'exico, D.F., 04510, 
M\'exico} 
\date{}
\maketitle
\begin{abstract}
A symmetry of the parameter space of interacting
boson models IBM-1 and IBM-2 is studied. 
The symmetry is associated  with  linear canonical transformations 
of boson operators, or, equivalently, with the existence of different 
realizations of the symmetry algebras 
of the models. The relevance of the parameter symmetry to physical 
observables is discussed. 
\end{abstract}
\section{Introduction}
It has been established recently~\cite{Izv,Goslar,PLB} that the 
Hamiltonian of the simplest version of the interacting boson model
IBM-1~\cite{Iachello} possesses an additional symmetry, the so-called 
parameter symmetry, that is a symmetry of the parameter space of the
model. The symmetry manifests itself in the existence of two sets of the
Hamiltonian parameters that generate 
identical spectra.  

The IBM-1 Hamiltonian has an algebraic structure 
characterized by the U(6) algebra. The spectrum and the eigenfunctions 
can be found analytically in three particular cases (the U(5), SU(3) and
SO(6) dynamical symmetry (DS)
limits).  A non-trivial issue of the parameter symmetry is that it
establishes an equivalence between the exactly solvable IBM-1 DS limits 
and transitional IBM-1 Hamiltonians of a general form. 
\par
The parameter symmetry is associated with canonical transformations 
of boson operators
linking different realizations  of SU(3) and SO(6) subalgebras in the
U(6) algebra~\cite{Goslar,PLB}. 
\par
In this paper we propose a generalization of the parameter symmetry 
concept on the case of IBM-2, the proton-neutron version of IBM. After a 
survey of the parameter symmetry of IBM-1, we turn the discussion to the 
structure of the general IBM-2 Hamiltonian and derive the
IBM-2 parameter symmetry relations.
\section{Parameter Symmetry of IBM-1}
Within IBM-1 nuclear states are labelled
by a fixed total number $N$ of bosons of two types, $s$ and $d$,
with quantum numbers $l^{\pi }=0^{+}$ and $l^{\pi }=2^{+}$,
respectively~\cite{Iachello}. The U(6) algebra is  generated by
36 bilinear combinations of boson operators: $s^{+}s$, 
$d^{+}_{\mu }d_{\nu}$, $d^{+}_{\mu }s$, $s^{+}d_{\mu}$.
The general 
IBM-1 Hamiltonian can be expressed as~\cite{Iachello}
\begin{equation}
\label{e}
\begin{array}{rl}
H(\{k_i\})\;=& H_0+k_1C_1[\mbox{U(5)}]+k_2C_2[\mbox{U(5)}]
+k_3C_2[\mbox{SO(5)}] \\
&+\: k_4C_2[\mbox{SO(3)}]+k_5C_2[\mbox{SO(6)}]+k_6C_2[\mbox{SU(3)}] \; ,
\end{array}
\end{equation}
where $C_1$ and $C_2$ stand for the first and the second rank
Casimir invariants of the algebras entering the reduction chains of the
U(6) algebra: 
\begin{equation}
\label{d}
{\renewcommand{\arraystretch}{0.1}
\begin{array}{cclc}
& & \mbox{U(5)}\supset \mbox{SO(5)}\supset
\mbox{SO(3)} & \qquad \mbox{I} \\
& \nearrow &  \\
\mbox{U(6)} & \rightarrow & \mbox{SU(3)}\supset \mbox{SO(3)}
& \qquad \mbox{II} \\
& \searrow &  \\
& & \mbox{SO(6)}\supset \mbox{SO(5)}\supset \mbox{SO(3)} \; 
& \qquad  \mbox{III} \\
\end{array} }
\end{equation}
We define the Casimir operators as in the book~\cite{Frank}:
\begin{equation}
\label{fg}
\begin{array}{@{}ll}
C_1[\mbox{U(5)}]=n_d \; , 
&
C_2[\mbox{SO(5)}]=2(T^3{\cdot }T^3)+2(T^1\cdot T^1) \; , 
\\
C_2[\mbox{U(5)}]=n_d(n_d{+}4) \; , 
&
C_2[\mbox{SO(6)}]=N(N{+}4)-4\left(P^{+}\cdot P\right) \; , 
\\ 
C_2[\mbox{SO(3)}]=10(T^1\cdot T^1) \; , 
&
C_2[\mbox{SU(3)}]=2(Q\cdot Q)+\frac{15}{2}(T^1\cdot T^1) \; ,
\end{array}
\end{equation}
where generators of the groups entering reduction chains (\ref{d}) are 
given in Table~1,
\begin{table}
\caption{Generators of U(6), U(5), SU(3), SO(6), SO(5) and SO(3) algebras}

\vspace{3mm}

\begin{center}
\begin{tabular}{|c|l|}
\hline
$\vphantom{\widetilde{\widetilde{\overline{\mbox{SO}}}}_{\pi \nu }}$    
Algebra & Generators \\ 
\hline
\hline
$\vphantom{\widetilde{\widetilde{\overline{\mbox{SO}}}}_{\pi \nu }}$    
U(6) &$[d^+{\times}s]^{(2)}_{\mu }$,
$[s^+{\times}\tilde d]^{(2)}_{\mu } $,  $S=[s^+{\times}s]^{(0)}_0$,
$T^\lambda_\mu=[d^+{\times}\tilde d]^{(\lambda )}_{\mu }$, \ 
$\lambda =0,1,2,3,4\!$ 
\\ \hline  
$\vphantom{\widetilde{\widetilde{\overline{\mbox{SO}}}}_{\pi \nu }}$    
U(5) & $T^\lambda_\mu=[d^+{\times}\tilde d]^{(\lambda )}_{\mu }$, \ \ 
$\lambda =0,1,2,3,4$ \\ \hline  
SU(3) & $Q_\mu =[d^+{\times}s+s^+{\times}\tilde d]^{(2)}_{\mu }
{-}\frac{\sqrt{7}}{2}[d^+{\times}\tilde d]^{(2)}_{\mu }$, \ \
$T^1_\mu=[d^+{\times}\tilde d]^{(1)}_{\mu }$ \\
\hline
SO(6) & $Q^0_\mu =[d^+{\times}s+s^+{\times}\tilde d]^{(2)}_{\mu }$, \ \
$T^\lambda_\mu=[d^+{\times}\tilde d]^{(\lambda )}_{\mu }$, \ \ 
$\lambda =1,3$ \\ \hline
SO(5) & $T^\lambda_\mu=[d^+{\times}\tilde d]^{(\lambda )}_{\mu }$, \ \ 
$\lambda =1,3$ \\
\hline
SO(3) & $T^1_\mu=[d^+{\times}\tilde d]^{(1)}_{\mu }$ \\
\hline
\hline                      
\end{tabular}
\end{center}
\end{table}
\begin{equation}
\label{def}
\begin{array}{ll}
n_d=(d^+ \cdot \tilde d) \; , 
& P=\frac12\left((\tilde d \cdot \tilde d)-(s\cdot s)\right) \;, 
\end{array}
\end{equation}                                     
$\tilde d_{\mu }=(-1)^{\mu }d_{-\mu }$, ${\left( t\cdot u\right) }$ and 
${\left[ t\times u\right] _{\mu }^{(\lambda )}}$ are 
scalar and  tensor products, respectively,  of spherical tensors $t$ 
and $u$. 
\par
Dynamical symmetry limits correspond to the cases when
the  Hamiltonian involves Casimir operators 
belonging to one of the reduction chains (\ref{d})  only, and hence
the eigenvalues and eigenfunctions can be found 
analytically. The spectrum of the IBM Hamiltonian in the case of a DS 
limit is one of the  typical nuclear
spectra~\cite{Iachello}:  vibrational in the U(5) 
DS limit ($k_5{=}k_6{=}0$), rotational in the SU(3) DS
limit ($k_1{=}k_2$${=}k_3$${=}k_5{=}$ 0) and $\gamma $-unstable in 
the SO(6) DS limit ($k_1{=}k_2{=}k_6{=}0$). A transitional
nuclear Hamiltonian that does not possess any DS, is
conventionally believed to generate a spectrum different from those
corresponding to any of the DS limits.
\par
As we have shown in Refs.~\cite{Izv,Goslar,PLB}, the IBM-1 Hamiltonian 
possesses a parameter symmetry, namely: 
\par
{\it Hamiltonians $H(\{k_i\})$ and $H(\{k^{\prime }_i\})$ defined by
Eq.}~(\ref{e}) {\it have identical spectra of eigenvalues
if the corresponding parameter sets $\{k_i\}$ and
$\{k^{\prime }_i\}$ are related as
\begin{equation}
\label{i1}
\begin{array}{l}
H^{\prime }_0=H_0,\ k^{\prime }_1=k_1+2k_6,\ k^{\prime }_2=k_2+2k_6,\
k^{\prime }_3=k_3-6k_6,\\
k^{\prime }_4=k_4+2k_6,\ k^{\prime }_5=k_5+4k_6,\ k^{\prime }_6=-k_6
\end{array}
\end{equation}
in the case $k_6\ne 0$, or as
\begin{equation}
\label{i2}
\begin{array}{l}
H^{\prime }_0=H_0+10k_5N,\ k^{\prime }_1=k_1+4k_5(N{+}2),\
 k^{\prime }_2=k_2-4k_5,\\
 k^{\prime }_3=k_3+2k_5,\ k^{\prime }_4=k_4,\
 k^{\prime }_5=-k_5,\ k^{\prime }_6=0
\end{array}
\end{equation}
in the case $k_6= 0$}.
\par
This statement was proved 
(see Refs.~\cite{Izv,Goslar,PLB} for details)
by constructing a unitary transformation 
$U$ such that 
$H(\{k_i^{\prime }\}) =UH(\{k_i\})U^{-1}$.
 \par
Thus, for any set of the IBM-1 parameters there is another set which 
generates the identical spectrum. The only exception is the U(5) DS limit 
when the two  sets of the parameters coincide as is seen from (\ref{i2}). 
\par
One of the most intriguing issues of the parameter symmetry is that it 
establishes the equivalence of 
the nuclear spectrum corresponding to a certain DS to 
the spectrum of a transitional IBM-1 Hamiltonian. As follows from
Eqs.~(\ref{i1}), the rotational spectrum of the SU(3) DS limit
($k_1{=}k_2{=}k_3{=}k_5{=}0$) appears to be equivalent to the 
spectrum of
the transitional Hamiltonian with the set of
parameters $\{k^{\prime }_i\}\equiv \{k^{\prime }_1{=}2k_6$,
$k^{\prime }_2{=}2k_6$, $k^{\prime }_3{=}{-}6k_6$,
$k^{\prime }_4{=}k_4{+}2k_6$, $k^{\prime }_5{=}2k_6$,
$k^{\prime }_6{=}{-}k_6\}$
that does not correspond to any DS. Similarly, it follows
from (\ref{i2}), that the $\gamma $-unstable spectrum of the SO(6) DS limit
($k_1{=}k_2{=}k_6{=}0$) can be obtained with the set of parameters
$\{k^{\prime }_i\}\equiv\{ k^{\prime }_6{=}0$,
$k^{\prime}_1{=}8(N{+}2)k_5$, $k^{\prime }_2{=}{-}8k_5$, 
$k^{\prime}_3{=}k_3{+}4k_5$,
$k^{\prime }_4{=}k_4$, $k^{\prime }_5{=}{-}k_5\}$
corresponding to the \mbox{U(5)--SO(6)} transitional nuclear
spectrum. In Ref.~\cite{Kusnezov}, such transitional Hamiltonians were 
referred to as the ones possessing  a hidden symmetry. 
\par
To reveal the origin of the parameter symmetry, we note 
that there is an ambiguity in definition of boson operators within 
IBM~\cite{PLB,Iachello,Frank,VanIsacker,discrete}. One
can apply to the boson operators gauge transformations
$R_s(\varphi_s)$ and $R_d(\varphi_d)$ defined as \cite{VanIsacker,Iachello} 
\begin{equation}
\label{rotation2R}
\begin{array}{ll}
R_s(\varphi_s)\,s^{+}=\exp(\mbox{i}\varphi _s/2)\,s^{+} \; , \qquad & 
R_s(\varphi_s)\,s=\exp(-\mbox{i}\varphi _s/2)\,s \; , \\
R_d(\varphi_d)\,d^{+}_{\mu }=\exp(\mbox{i}\varphi _d/2)\,d^{+}_{\mu } \; ,
\qquad  & 
R_d(\varphi_d)\,\tilde d_{\mu }=\exp(-\mbox{i}\varphi_d/2)\,\tilde d_{\mu } \; .
\end{array}
\end{equation}
Note that these transformations are the canonical ones, i.e.\ they do
not violate the boson commutation relations. 
However, the structure of the IBM Hamiltonian implies severe
restrictions on the use of the  transformations (\ref{rotation2R}). For
example, in the case of the general IBM Hamiltonian, one can only
apply to the Hamiltonian the gauge transformation 
$R(\varphi)\equiv R_s(\varphi_s)\times R_d(\varphi_d)$ with
$\varphi\equiv(\varphi_s-\varphi_d)/2=0,\,\pi$ and arbitrary
$\tilde\varphi\equiv(\varphi_s+\varphi_d)/2$~\cite{VanIsacker,discrete}. 
Similarly, in the case of the 
transitional SO(6)--U(5) IBM Hamiltonian with $k_{6}=0$, one can use
the gauge transformation $R(\varphi)$  with
$\varphi=0,\,\displaystyle\frac{\pi}{2},\,\pi,\,\frac{3\pi}{2}$.
One can also apply to the boson operators the 
 particle-hole conjugation  $\tilde R$~\cite{Dieperink,Frank,discrete}
defined as
\begin{equation}
\label{ph2R}
\begin{array}{ll}
\tilde{R}\,s^{+}=s \; , \qquad & \tilde{R}\,s=-s^{+} \; , \\
\tilde{R}\, d^{+}_{\mu }=\tilde d_{\mu } \; , \qquad & 
\tilde{R}\,\tilde d_{\mu} =-d^{+}_{\mu } \; ,
\end{array}
\end{equation}  
and operators $\tilde{R}(\varphi)\equiv \tilde{R}\times R(\varphi)$ that
are consistent with the Hamiltonian structure provided that 
$\varphi=0,\,\displaystyle\frac{\pi}{2},\,\pi,\,\frac{3\pi}{2}$ in the case 
$k_{6}=0$ or $\varphi=0,\,\pi$ in the case $k_{6}\ne 0$. The operators
${R}(\varphi)$ and $\tilde{R}(\varphi)$ comprise a point group studied
elsewhere~\cite{discrete}. 

We use the following notations for operators subjected to the
transformations $R(\varphi)$ and $\tilde R(\varphi)$:
${^{\alpha}O}\equiv R(\varphi)\,O$  and 
${^{-\alpha}O}\equiv \tilde R(\varphi)\,O$, $\alpha=\varphi/\pi$.

Hamiltonians ${^{\alpha}H}(\{k_i\})\equiv  R(\varphi)\,H(\{k_i\})$ are,
of course, isospectral with the initial Hamiltonian  $H(\{k_i\})$ 
[Hamiltonians 
${^{-\alpha}H}(\{k_i\})\equiv \tilde R(\varphi)\,H(\{k_i\})$ may
be not isospectral with  $H(\{k_i\})$; however, one can always find a
simple and not very restrictive constraint on the parameters $k_i$
that will guarantee the 
isospectrality of ${^{-\alpha}H}(\{k_i\})$ and
$H(\{k_i\})$]. Thus we can use transformations (\ref{rotation2R}) [and
in some cases (\ref{ph2R})] to study parameter symmetries and hidden
symmetries of IBM.

For example, the Hamiltonian 
${^{1}H}(\{k_i\})\equiv  R(\pi)\,H(\{k_i\})$ is isospectral but not
equivalent to the initial Hamiltonian  $H(\{k_i\})$.  Using Table~1 and
expressions (\ref{e}), (\ref{fg}), (\ref{def}) and  (\ref{rotation2R}),
one can obtain \cite{PLB} that ${^{1}H}(\{k_i\})=H(\{k^{\prime}_i\})$ where the
set of parameters $\{k^{\prime}_i\}$ is defined by the parameter
symmetry relation (\ref{i1}). Thus, the transformation $R(\pi)$
is equivalent to  the parameter symmetry transformation (\ref{i1}).

Note that, as is seen from  Eqs.~(\ref{fg})--(\ref{def}) and Table~1, 
only the Casimir operator $C_{2}[\mbox{SU(3)}]$  and the SU(3) generator
$Q_\mu$ are changed under the transformation $R(\pi)$: 
\begin{equation}
R(\pi)\,Q_\mu=
{^{1}Q}_{\mu }= -[d^{+}{\times }\,s\,+\,s^{+}{\times }\,
\tilde{d}]^{(2)}_{\mu }
-\frac{\sqrt{7}}{2}[d^{+}{\times }\,\tilde{d}]^{(2)}_{\mu }\; . \label{Qbar}
\end{equation}
The quadrupole operators $Q_{\mu }$ and ${^{1}Q}_{\mu }$ correspond to 
different embeddings of the SU(3) subalgebra in the U(6) algebra 
[see also \cite{Dieperink} for other realizations of SU(3)]. Using
parameter symmetry transformation (\ref{i1}) it is easy to express
the Casimir operator $C_{2}\!\left[{\rm {SU^{1}(3)}}\right] $ of 
the ${\rm SU}^{1}(3)$ algebra
associated with the quadrupole operator (\ref{Qbar}) through 
$C_{2}\left[\mbox{SU(3)}\right]$ and  Casimir operators of other
algebras~\cite{Goslar,PLB}:
\begin{equation}
\begin{array}{rl}
C_{2}\!\left[{\rm{SU^{1}(3)}}\right] = & 2C_1[\mbox{U(5)}]
+2C_{2}[\mbox{U(5)}]-6C_{2}[\mbox{SO(5)}] \\ 
& +\ 2C_{2}[\mbox{SO(3)}]+4C_{2}[\mbox{SO(6)}]-C_{2}[\mbox{SU(3)}] \; .
\end{array}
\label{m1}
\end{equation}
\par
In the case $k_{6}=0$ we have ${^{1}H}(\{k_i\})=H(\{k_i\})$ and 
the transformation $R(\pi)$ does not generate the parameter symmetry.
However, in this case we can apply the transformation $R(\pi/2)$ to the 
Hamiltonian. 
Using Table~1 and expressions (\ref{e}), (\ref{fg}), (\ref{def}) and 
(\ref{rotation2R}),
we obtain \cite{PLB} that $^{\slantfrac{1}{2}}H(\{k_i\})=H(\{k^{\prime}_i\})$
where the 
set of parameters $\{k^{\prime}_i\}$ is defined by the parameter
symmetry relation (\ref{i2}). Hence, the transformation $R(\pi/2)$
is equivalent to  the parameter symmetry transformation (\ref{i2}).

With the help of the transformation $R(\pi/2)$ we obtain a new
monopole operator:
\begin{equation}
\overline{P}\equiv {^{\slantfrac{1}{2}}P}=R(\pi/2)\, P=
\frac{1}{2}\left((\tilde d{\cdot } \tilde d)+
(s{\cdot } s)\right) .
\label{Pbar}
\end{equation}
This monopole operator corresponds to an alternative embedding of the
SO(6) subalgebra in the U(6) algebra \cite{VanIsacker,Frank}. 
Using the parameter symmetry relation (\ref{i2}) it is easy to
obtain~\cite{Goslar,PLB} the following expression for 
the Casimir operator of the $\overline{\mbox{SO(6)}}$ algebra associated with 
the monopole operator $\overline{P}$: 
\begin{equation}
C_{2}\!\left[ \overline{\mbox{SO(6)}}\right] =
10N{+}4(N{+}2)\,C_{1}[\mbox{U(5)}]
{-}4C_{2}[\mbox{U(5)}]{+}2C_{2}[\mbox{SO(5)}]{-}C_{2}[\mbox{SO(6)}]\,.  
\label{n1}
\end{equation}
\par
Note, that the Casimir operators of alternatively embedded 
algebras SU$^1$(3) and $\overline{\rm SO}(6)$ are not independent from 
the Casimir operators of  other algebras and  should not be
included into the general Hamiltonian (\ref{e}). 
\par
The transformations $\tilde R(\varphi)$ do not generate new parameter
symmetries. However one more parameter symmetry relation can be
obtained in the case of IBM-1 that is not associated with the
transformations $R(\varphi)$ and  $\tilde R(\varphi)$ (see~\cite{PLB}
for a more detailed discussion).  
%
%
\par
Usually in applications the Hamiltonian parameters $\{k_i\}$ are obtained
by the fit to nuclear spectra. Due to the parameter symmetry, the fit
of the parameters appears to be ambiguous. 
To discriminate between the two sets of parameters
giving rise to identical spectra,
it is natural to study electromagnetic
transitions. 

In the consistent-$Q$ formalism (C$Q$F)~\cite{CQF}, both
monopole-monopole  
(${P^{+}\cdot P}$) and quadru\-pole-quadru\-pole 
($Q\cdot Q$) interactions  are replaced in the Hamiltonian by a single
term ($Q^{\chi }\cdot Q^{\chi }$) where
the generalized quadrupole operator
\begin{equation}
\label{QHI}
Q^{\chi }_{\mu }=[d^{+}{\times}\,s +s^{+}{\times}\,\tilde d]^{(2)}_{\mu }
+\chi\,[d^{+}{\times}\, \tilde d]^{(2)}_{\mu }.
\end{equation}
Operator $Q^{\chi }$ is used for calculations of
$E2$-transition rates within C$Q$F.
Applying transformation $R(\pi)$ to the Hamiltonian
$H(\{k_i\})$, we find out that the only term in the new Hamiltonian
${^{1}H}(\{k_i\})=R(\pi)\,H(\{k_i\})$ that differs from the corresponding
term in the initial Hamiltonian $H(\{k^{\prime}_i\})$, is the
generalized quadrupole-quadrupole interaction: 
\begin{equation}
\label{1Q1Q}
\left({^{1}Q}^{\chi }{\cdot }\,{^{1}Q}^{\chi }\right)=R(\pi)\, 
\left({\vphantom{^{1}}Q^{\chi }}{\cdot }\,{Q^{\chi }}\right),
\end{equation} 
where 
\begin{equation}
\label{1Qchi}
{^{1}Q}^{\chi}_{\mu }\equiv R(\pi)\,{Q^{\chi }_{\mu }}=-Q^{-\chi }_{\mu } \; .
\end{equation}
%
The consistent transformation of the $E2$ transition operator
(\ref{QHI}) according to (\ref{1Qchi}) and of the generalized
quadrupole-quadrupole interaction in the Hamiltonian according to
(\ref{1Q1Q}), guarantees that the $E2$ transition rates remain unchanged.
Therefore in the general case $\chi\ne 0$ that corresponds to 
$k_{6}\neq 0$,  the $E2$ transition rates cannot be used to distinguish between
two sets of Hamiltonian parameters $\{k_i\}$ and $\{k'_i\}$ related by
the parameter symmetry (\ref{i1}), at least within the C$Q$F formalism.
if it is believed that the
C$Q$F ansatz is an adequate prescription for the electromagnetic
transition operator. 
We note that in the general case $\chi\ne 0$  
the type of the generalized quadrupole-quadrupole interaction [whether
it is of the form $(Q^{\chi}{\cdot }\,Q^{\chi })$ or 
$(^1Q^{\chi}{\cdot }\,^1Q^{\chi })$]
is unambiguously determined by the set
of the Hamiltonian parameters $\{k_{i}\}$.  We have shown in \cite{PLB} that
this is due to the fact that the generalized
quadrupole-quadrupole interaction  
includes the monopole-monopole term (${P^{+}\cdot P}$). 

In the case  $\chi =0$ ($k_6=0$), we apply the transformation
$R(\pi/2)$ to the Hamiltonian and to the quadrupole operator
$Q^0_\mu$ to obtain
\begin{equation}
\label{Q0bar}
\overline{Q^{0}_{\mu }}\equiv ^{\slantfrac{1}{2}\!\!}Q^{0}_\mu\equiv
R(\pi/2)\,Q^{0}_{\mu } =-\mbox{i}[d^{+}{\times }\,s\ 
-\ s^{+}{\times }\,\tilde d]^{(2)}_{\mu }\, .
\end{equation}
However, in this case the generalized quadrupole-quadrupole
interaction is ambiguous. As we have shown in Ref.~\cite{PLB}, the
parameter symmetry relation (\ref{i2}) can be used to derive
\begin{equation}
\left(\vphantom{\overline{Q^0}}Q^{0}{\cdot }\,Q^{0}\right)=
-\left( \overline{Q^{0}}{\cdot }\,\overline{Q^{0}}
\right){+}10N{+}4(N{+}2)\,C_{1}[\mbox{U(5)}]{-}4C_{2}[\mbox{U(5)}]
{-}2C_{2}[\mbox{SO(5)}]\,.  \label{Q0ambig}
\end{equation}
Thus in the case $k_{6}=0$ the IBM-1 Hamiltonian can be expressed either
through $\left(\vphantom{\overline{Q^0}} Q^{0}{\cdot }\,Q^{0}\right)$
or alternatively through
$\left( \overline{Q^{0}}{\cdot }\,\overline{Q^{0}}\right) $. 
As a result, the definition of the $E2$ transition operator appears to be 
ambiguous. Due to this ambiguity, the electromagnetic transition rates
cannot be used to distinguish among two sets of Hamiltonian parameters
$\{k_i\}$ and $\{k'_i\}$ related by the parameter symmetry (\ref{i2}). 
The origin of the ambiguity of the  generalized quadrupole-quadrupole
interaction is that the  quadrupole-quadrupole interaction 
$\left( Q{\cdot }\,Q\right)$ is not present in the Hamiltonian in the
case $k_6=0$ and the operators 
$\left(\vphantom{\overline{Q^0}} Q^{0}{\cdot }\,Q^{0}\right)$ and 
$\left( \overline{Q^{0}}{\cdot }\,\overline{Q^{0}}\right)$  within 
C$Q$F replace  the monopole-monopole term (${P^{+}\cdot P}$) in the
Hamiltonian  (see Ref.~\cite{PLB} for more details).
\par
There is another possibility to distinguish among the two Hamiltonians
related by the parameter 
symmetry  in the case $k_{6}=0$. This possibility
stems from  the $N$-dependence of the parameter symmetry~(\ref{i2}).  
Since the relations~(\ref{i2}) involve the total
number of bosons $N$, the two sets of the parameters can generate
identical spectra for some particular nucleus only, the predictions
for the spectra of its isotopes or
isotones should differ. It is conventionally supposed 
(see for example Ref.~\cite{Ndependence}) that the
spectra of neighboring even-even nuclei are described by the same set
of the IBM parameters, hence one can
discriminate between the parameter sets $\{k_i\}$ and $\{k'_i\}$ related 
according to (\ref{i2}) by comparing the
spectra of the neighboring nuclei. 

This is illustrated by Fig.~1 where the spectra of three Pt isotopes
are presented. 
\begin{figure}
\centerline{\epsfig{file=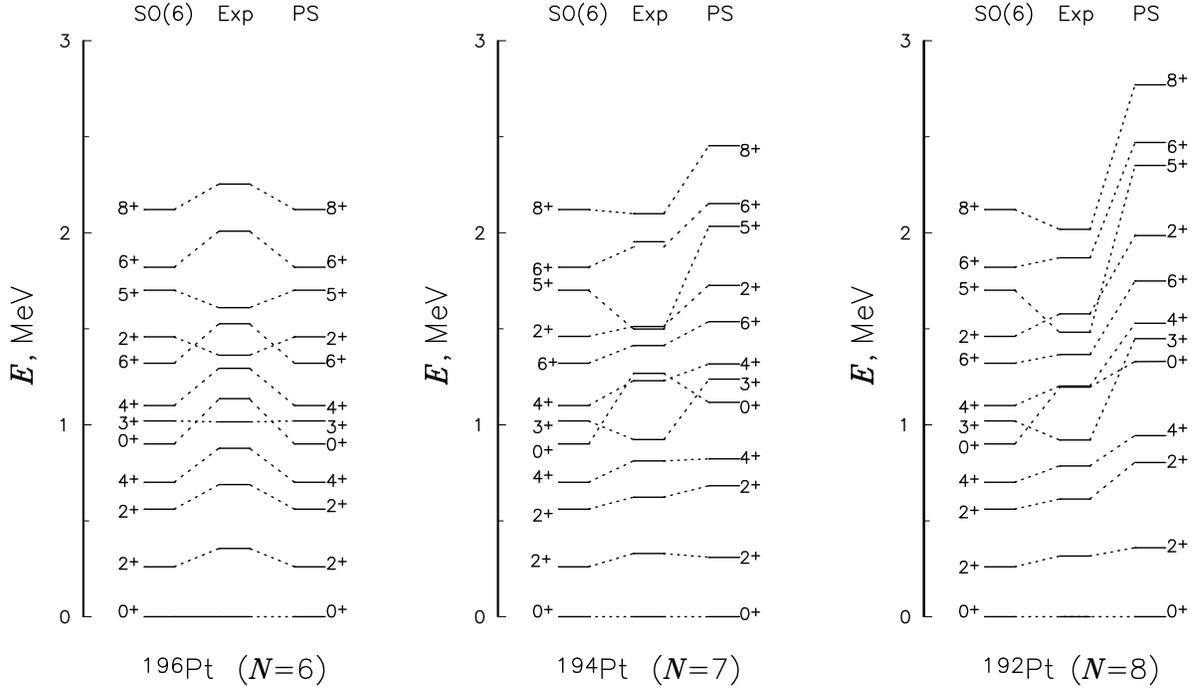,width=0.98\textwidth}}
\caption{Few lowest levels of each $J^{\pi}$ of Pt isotopes. SO(6):
calculations within SO(6) DS limit with the parameters
suggested in Ref.~\protect\cite{Iachello}; PS: calculations with the
set of parameters obtained using 
(\protect\ref{i2}) with $N=6$; Exp: experimental data of
Ref.~\protect\cite{ExpPt}.}
\end{figure}
The set of
parameters $k_{1}=k_{2}=k_{6}=0$, $k_{3}=50$~keV, $k_{4}=10$~keV, $
k_{5}=-42.75$~keV was suggested in \cite{Iachello}
for the description of $^{196}$Pt ($N$=6) within the SO(6) DS 
limit of IBM.
The corresponding spectra are given in the left columns labelled by SO(6).
The set of parameters $k_{1}^{\prime }=-1368$~keV, $k_{2}^{\prime }=171$~keV, 
$k_{3}^{\prime }=-35.5$~keV, $k_{4}^{\prime }=10$~keV, $k_{5}^{\prime }
=42.75$~keV and $k_{6}^{\prime }=0$ is obtained using (\ref{i2}) with $N=6$.
The corresponding spectra are given in the right columns labelled by PS.
The SO(6) and parameter symmetry spectra are, of course, identical in the 
case of $^{196}$Pt but differ for other Pt isotopes. 
\par
As is seen from (\ref{e}), (\ref{fg}) and (\ref{def}),
the transformation $R(\pi/2)$ or, equivalently, the parameter symmetry
transformation (\ref{i2}), changes  the sign of the monopole-monopole
interaction $(P^{+}{\cdot }\,P)$ in the Hamiltonian. This sign change
manifests itself in the spectra of neighboring nuclei. It is usually
supposed that the pairing (monopole-monopole) interaction should be
attractive, i.e. $k_5<0$. Note that the set of parameters suggested in
Ref.~\cite{Iachello} with attractive pairing interaction fitted to
$^{196}$Pt, reproduces the  experimental data on $^{192}$Pt and
$^{194}$Pt better (see Fig.~1) than the other set with 
$k^{\prime }_5>0$. 

The generalized
quadrupole-quadrupole interaction $(Q^{\chi }{\cdot }\,Q^{\chi })$
incorporates both quadrupole-quadrupole $(Q{\cdot }\,Q)$ and
pairing  $(P^{+}{\cdot }\,P)$ interactions. The transformation
$R(\pi)$, or, equivalently, the parameter symmetry transformation
(\ref{i1}) changes  only the sign of the quadrupole-quadrupole
interaction $(Q{\cdot }\,Q)$ in the Hamiltonian, as is seen from 
Table~1 and Eqs.~(\ref{e}), (\ref{fg}) and (\ref{def}); the 
monopole-monopole and other multipole-multipole terms are not effected
by the transformation $R(\pi)$. Contrary to that of the pairing
interaction, the sign of the quadrupole-quadrupole interaction
$(Q{\cdot }\,Q)$ is of no physical importance and is indistinguishable
in applications as we have shown above. 


\section{ Parameter Symmetry of IBM-2}

IBM-2~\cite{IBM2_1,Iachello} is a proton-neutron version of
IBM. Within this model, $s$ and $d$ bosons are introduced
in the proton and neutron subspaces independently.
The symmetry algebra of the model is
U$_{\pi }$(6)$\otimes $U$_{\nu }$(6) generated by 72 bilinear operators 
$s_{\rho }^{+}s_{\rho }$, $d^{+}_{\rho \mu }d_{\rho \nu}$, 
$d^{+}_{\rho \mu }s_{\rho }$, $s^{+}_{\rho }d_{\rho \mu}$
($\rho= \pi $, $\nu $).

The general IBM-2 Hamiltonian $H$ consists of
proton part $H_{\pi }$, neutron part $H_{\nu }$ and 
proton-neutron interaction $V_{\pi \nu }\,$,
\begin{equation}
\label{Hibm2}
H\left(\{k_i^\pi,k_i^\nu,k_i\}\right)=
H_{\pi} \left(\{k_i^\pi\}\right)
+H_{\nu}\left(\{k_i^\nu\}\right) +V_{\pi \nu }\left(\{k_i\}\right) ,
\end{equation}
and is characterized by 21 independent parameters
$\{k_i^\pi,k_i^\nu,k_i\}$~\cite{Iachello}. 
The proton 
and neutron
parts of the Hamiltonian,
$H_{\rho}\left(\{k_i^\rho\}\right)$, $\rho =\pi, \nu$, 
are just the IBM-1 Hamiltonians and are given by  (\ref{e}) with $H_0=0$.
It is desirable to express the proton-neutron interaction 
$V_{\pi \nu}\left(\{k_i\}\right)$ as a superposition of Casimir operators of
combined  proton-neutron subalgebras G$_{\pi \nu}$
of the U$_{\pi }$(6)$\otimes $U$_{\nu }$(6) algebra that
enter the reduction chains starting with
U$_{\pi }$(6)$\otimes $U$_{\nu }$(6) and ending with SO$_{\pi \nu }$(3).
The generators ${\cal G}_{\pi \nu }$ of the combined proton-neutron
algebras G$_{\pi \nu}$ are of  the form  
${\cal G}_{\pi \nu }={\cal G}_{\pi }+{\cal G}_{\nu }$ where 
${\cal G}_{\pi }$ and ${\cal G}_{\nu }$ are generators of the
corresponding proton and neutron algebras, respectively. For example,
the generators of the  SO$_{\pi \nu}$(6) algebra are
$Q^0_{\pi \mu }+Q^0_{\nu \mu }\,$, 
$T^1_{\pi \mu }+T^1_{\nu \mu }\,$, and $T^3_{\pi \mu }+T^3_{\nu \mu }\,$.
                        
The U$_{\pi }$(6)$\otimes $U$_{\nu }$(6) algebra has a number of
appropriate reduction chains.
There are three  types of the reduction chains which include U(5),
SU(3) and SO(6) subalgebras~\cite{Frank}:
 
1. U(5) DS chains:
\begin{equation}
\label{chU5}
{\renewcommand{\arraystretch}{0.0}
\begin{array}{ccccccccc}
\mbox{U$_{\pi }$(6)} & \rightarrow & \mbox{U$_{\pi }$(5)} & 
\rightarrow & \mbox{SO$_{\pi }$(5)} & \rightarrow & 
\mbox{SO$_{\pi }$(3)} & & \\
& \searrow & & \searrow & & \searrow & & \searrow &  \\
& & \mbox{U}_{\pi \nu }(6) & \rightarrow &
\mbox{U}_{\pi \nu }(5) & \rightarrow & \mbox{SO}_{\pi \nu }(5) & 
\rightarrow & \mbox{SO}_{\pi \nu }(3)  \\
& \nearrow & & \nearrow & & \nearrow & & \nearrow &  \\
\mbox{U}_{\nu }(6) & \rightarrow & \mbox{U}_{\nu }(5) & 
\rightarrow & \mbox{SO}_{\nu }(5) & \rightarrow & 
\mbox{SO}_{\nu }(3) & & 
\end{array} }
\end{equation}

2.  SU(3) DS chains:
\begin{equation}
\label{chSU3}
{\renewcommand{\arraystretch}{0.0}
\begin{array}{ccccccc}
\mbox{U$_{\pi }$(6)} & \rightarrow & \mbox{SU$_{\pi }$(3)} & 
\rightarrow & \mbox{SO$_{\pi }$(3)} & & \\
& \searrow & & \searrow & & \searrow &  \\
& & \mbox{U}_{\pi \nu }(6) & \rightarrow &
\mbox{SU$_{\pi \nu }$(3)} & \rightarrow & \mbox{SO$_{\pi \nu }$(3)}  \\
& \nearrow & & \nearrow & & \nearrow &  \\
\mbox{U$_{\nu }$(6)} & \rightarrow & \mbox{SU$_{\nu }$(3)} & 
\rightarrow & \mbox{SO$_{\nu }$(3)} & & 
\end{array} }
\end{equation}

3. SO(6) DS chains:
\begin{equation}
\label{chSO6}
{\renewcommand{\arraystretch}{0.0}
\begin{array}{ccccccccc}
\mbox{U}_{\pi }(6) & \rightarrow & \mbox{SO}_{\pi }(6) & \rightarrow & 
\mbox{SO}_{\pi }(5) & \rightarrow & \mbox{SO}_{\pi }(3) & & \\
& \searrow & & \searrow & & \searrow & & \searrow &  \\
& & \mbox{U}_{\pi \nu }(6) & \rightarrow &
\mbox{SO}_{\pi \nu }(6) & \rightarrow & 
\mbox{SO}_{\pi \nu }(5) & \rightarrow & \mbox{SO}_{\pi \nu }(3)  \\
& \nearrow & & \nearrow & & \nearrow & & \nearrow & \\
\mbox{U}_{\nu }(6) & \rightarrow & \mbox{SO}_{\nu }(6) & \rightarrow & 
\mbox{SO}_{\nu }(5) & \rightarrow & \mbox{SO}_{\nu }(3) & & 
\end{array} }
\end{equation}                

We note that the set of Casimir operators provided by the algebras
entering the reduction chains 
(\ref{chU5})--(\ref{chSO6}),  is not
complete enough to express the general IBM-2 Hamiltonian (\ref{Hibm2}).
The problem is partly solved by adding the $\overline{\mbox{SO}}(6)$
DS reduction chains~\cite{Frank} to the  reduction chains
(\ref{chU5})--(\ref{chSO6}):

4. $\overline{\mbox{SO}}(6)$ DS chains:
\begin{equation}
\label{chbarSO6}
{\renewcommand{\arraystretch}{0.0}
\begin{array}{ccccccccc}
\mbox{U}_{\pi }(6) & \rightarrow & \overline{\mbox{SO}}_{\pi }(6) 
& \rightarrow & 
\mbox{SO}_{\pi }(5) & \rightarrow & \mbox{SO}_{\pi }(3) & & \\
& \searrow & & \searrow & & \searrow & & \searrow &  \\
& & \mbox{U}_{\pi \nu }(6) & \rightarrow &
\overline{\mbox{SO}}_{\pi \nu }(6) & \rightarrow & 
\mbox{SO}_{\pi \nu }(5) & \rightarrow & \mbox{SO}_{\pi \nu }(3)  \\
& \nearrow & & \nearrow & & \nearrow & & \nearrow & \\
\mbox{U}_{\nu }(6) & \rightarrow & \overline{\mbox{SO}}_{\nu }(6) 
& \rightarrow & 
\mbox{SO}_{\nu }(5) & \rightarrow & \mbox{SO}_{\nu }(3) & & 
\end{array} }
\end{equation}

The reduction chains (\ref{chU5})--(\ref{chbarSO6}) will be referred
to as standard DS reduction chains.

Contrary to the  Casimir operators 
$C_2[\overline{\mbox{SO}}_{\pi}(6)]$ and 
$C_2[\overline{\mbox{SO}}_{\nu}(6)]$ [see Eq. (\ref{n1})],
the Casimir operator $C_2[\overline{\mbox{SO}}_{\pi \nu }(6)]$  is an
additional independent operator that can be used for the construction
of the general IBM-2 Hamiltonian.
However we still do not have a complete set of independent Casimir
operators. To obtain this set we should look for alternative
embeddings of the combined proton-neutron algebras. All the
alternative subalgebras can be obtained by applying all possible 
transformations $R_\rho(\varphi_\rho)$ and 
$\tilde R_\rho(\varphi_\rho)$ to the generators of all subalgebras in
the reduction chains  (\ref{chU5})--(\ref{chbarSO6}). As a result, we
obtain alternative subalgebras ${\rm G}_\rho^{\alpha_\rho}$ and
${\rm G}_{\pi\nu}^{\alpha_\pi\alpha_\nu}$ with generators
${^{\alpha_\rho}{\cal G}_\rho}$ and 
$\;{^{\alpha_\pi\alpha_\nu}{\cal G}_{\pi\nu}}
={^{\alpha_\pi}{\cal G}_\pi}+{^{\alpha_\nu}{\cal G}_\nu}$, respectively.
For example, the generators of the algebra 
SO$^{0 1}_{\pi \nu}$(6) are $Q^0_{\pi \mu }-Q^0_{\nu \mu }$, 
$T^1_{\pi \mu }+T^1_{\nu \mu }$ and $T^3_{\pi \mu }+T^3_{\nu \mu}$ [see
Eq.~(\ref{1Qchi}) for the expression of ${^{1}Q^0_{\nu\mu}}$]. The
$\overline{\mbox{SO}}_{\pi\nu}(6)$ algebra is the  
${\rm 
SO}_{\pi\nu}^{\!\!\!\!\!\!\!\slantfrac{1}{2}\!\!\!\!\!\!\!\slantfrac{1}{2}}(6)$
algebra in these notations. The  SU$^{0 -1}_{\pi \nu}$(3) algebra
is equivalent to the  
SU$^{*}_{\pi \nu }$(3) algebra introduced in Ref.~\cite{Dieperink} 
for the description of triaxial shapes within IBM-2.  

In such a way we obtain a large number of
alternative algebras. However, some of them are equivalent. For
example, any algebra ${\rm G}_{\pi\nu}^{\alpha_\pi\alpha_\nu}$ is equivalent
to its proton-neutron particle-hole counterpart
algebra ${\rm G}_{\pi\nu}^{-\alpha_\pi-\alpha_\nu}$ --- the
relative sign of $\alpha_\nu$ and $\alpha_\pi$ is only important, changing
the sign of both $\alpha_\nu$ and $\alpha_\pi$ we do not obtain a new
algebra. As follows from our analysis, there exist 2 different realizations of
U$_{\pi \nu }$(5), 2 different realizations of SO$_{\pi \nu }$(6),
2 different realizations of $\overline{\mbox{SO}}_{\pi \nu }$(6),  
and 8 different realizations of SU$_{\pi \nu }$(3).

The alternative algebras provide us with Casimir operators that can be
used for the construction of $V_{\pi \nu}\left(\{k_i\}\right)$,
however not all of these Casimir operators are independent. 
For example, the Casimir operators of alternative proton or neutron
algebras ${\rm G}_\rho^{\alpha_\rho}$ can be expressed through the Casimir
operators of untransformed algebras ${\rm G}_\rho$ [see Eqs.~(\ref{m1}) and
(\ref{n1})]; the rank-1 Casimir operator of any of 
${\rm U}^{\alpha_\pi\alpha_\nu}_{\pi\nu}(5)$ algebras can be expressed
through the Casimir operators of ${\rm U}^{\alpha_\pi}_{\pi}(5)$ and 
${\rm U}^{\alpha_\nu}_{\nu}(5)$: 
$C_1[{\rm U}^{\alpha_\pi\alpha_\nu}_{\pi\nu}(5)]=
C_1[{\rm U}^{\alpha_\pi}_{\pi}(5)]+
C_1[{\rm U}^{\alpha_\nu}_{\nu}(5)]$, etc. 
So, we should choose a set of independent rank-2 Casimir
operators of combined proton-neutron subalgebras. 
We suggest to include in this set the Casimir operators of
U$_{\pi \nu }$(6), U$_{\pi \nu }$(5), SO$_{\pi \nu }$(6),
$\overline{\mbox{SO}}_{\pi \nu }$(6), SU$_{\pi \nu }$(3), 
SU$^{01}_{\pi \nu }$(3) and SU$^{11}_{\pi \nu }$(3). 
The Casimir operators of all the rest proton-neutron algebras  can be 
expressed through the ones included in the set, e.g.,  
\begin{equation}
\label{CT9}
\begin{array}{rl}
C_2\left[\mbox{SU}^{0 -1}_{\pi \nu }(3)\right]\: 
 =&-\:C_2\,[\mbox{SU}^{01}_{\pi \nu }(3)]
+2C_2\left[\mbox{SU}_{\pi }(3)\right]
+2C_2\,[\mbox{SU}^1_{\nu }(3)] \\[2mm]
& \displaystyle
 +\:\frac32\left\{C_2\left[\mbox{SO}_{\pi \nu }(3)\right] 
-C_2\left[\mbox{SO}_{\pi }(3)\right]-C_2\left[\mbox{SO}_{\nu }(3)\right] 
\vphantom{^2}
\right\} .
\end{array}
\end{equation}
The Casimir operator $C_2[\mbox{U}(6)]$  not
defined above can be expressed as

\begin{equation}
\label{C2U6}
C_2\left[\mbox{U}(6)\right]=
(S\cdot S)+\frac12(Q^{0}\cdot Q^{0})
+\frac12(\overline{Q^{0}}\cdot \overline{Q^{0}})
+\sum_{\lambda =0}^{4}(T^{\lambda }\cdot T^{\lambda }) \; .
\end{equation}

The proton-neutron interaction  $V_{\pi \nu }\left(\{k_i\}\right)$ 
we express through the set of independent Casimir operators as
\begin{equation}
\label{Vpn}
\begin{array}{rl}
V_{\pi \nu }\left(\{k_i\}\right)\:
=&H_0+kC_2[\mbox{U}_{\pi \nu }(6)]
+k_2C_2[\mbox{U}_{\pi \nu }(5)]
+k_3C_2[\mbox{SO}_{\pi \nu }(5)] \\
&+\:k_4C_2[\mbox{SO}_{\pi \nu }(3)]
+k_5C_2[\mbox{SO}_{\pi \nu }(6)]
+k_6C_2[\overline{\mbox{SO}}_{\pi \nu }(6)] \\
&+\:k_7C_2[\mbox{SU}_{\pi \nu }(3)]                             
+k_8C_2[\mbox{SU}^{01}_{\pi \nu }(3)]
+k_9C_2[\mbox{SU}^{11}_{\pi \nu }(3)] \; .
\end{array}
\end{equation}
Note that the set  of independent Casimir operators is not unique and,
as a result, alternative expressions for 
$V_{\pi \nu }\left(\{k_i\}\right)$ can be suggested.
Another possible choice of the operators was used in Ref.~\cite{Leviatan2}.

It is seen that the construction of different realizations of boson
algebras plays  an important role in IBM-2.
The incompleteness of boson Hamiltonians in the form of
superposition of Casimir invariants of different groups determined by
standard DS reduction chains, is a common property of systems of
two (or more) independent subsystems, e.g. it is also a property of the
vibron model of 
triatomic molecules with the symmetry algebra 
U$_1$(4)$\otimes $U$_2$(4)~\cite{LeviatanVM}.

The standard reduction chains  (\ref{chU5})--(\ref{chbarSO6}) define
standard DS limits of IBM-2. As the Casimir operators of
SU$^{01}_{\pi\nu}$ and SU$^{11}_{\pi\nu}$ are present in the
Hamiltonian, we can also define non-standard SU$^{01}_{\pi\nu}$ and
SU$^{11}_{\pi\nu}$ DS limits of IBM-2 that are associated with non-standard
SU$^{01}$ and SU$^{11}$ DS reduction chains, respectively. The definition
of non-standard DS limits of the model is, of course, ambiguous
because of the ambiguity of definition of the complete set of Casimir
operators.  

Applying all possible 
transformations $R_\rho(\varphi_\rho)$ and 
$\tilde R_\rho(\varphi_\rho)$ to all subalgebras in
the standard reduction chains  (\ref{chU5})--(\ref{chbarSO6}), we obtain all
alternative reduction chains. Some of these reduction chains appear to
be equivalent to some of the others, some of them are equivalent to
some of the standard or non-standard DS reduction chains.
However the set of independent alternative
reduction chains can be easily defined. These  independent alternative
reduction chains give rise to hidden symmetries of the model.
It is interesting that some of the hidden symmetries may be obtained
by means of transformations (e.g., by particle-hole transformations)
that are not isospectral. 

Applying all possible 
transformations $R_\rho(\varphi_\rho)$ and 
$\tilde R_\rho(\varphi_\rho)$ to the general Hamiltonian (\ref{Hibm2}), 
we obtain a general IBM-2 Hamiltonian that can be
(i)~identical to the initial   Hamiltonian (\ref{Hibm2}),
(ii)~non-identical to but isospectral with  the initial   Hamiltonian
(\ref{Hibm2}), or (iii)~non-isospectral with  the initial   Hamiltonian
(\ref{Hibm2}). In the case (ii) we obtain standard parameter
symmetries of IBM-2 that are valid without restrictions on the
parameters of the model. The standard parameter symmetries are listed
in Table 2.
\begin{table}[t]
\caption{Standard parameter symmetry relations for the IBM-2 Hamiltonian}

\vspace{3mm}

\begin{center}
\begin{tabular}{|l|l|l|l|}
\cline{1-4}
A & B & C & D \\
\cline{1-4}  
$\vphantom{\widetilde{\widetilde{H}}}$    
 $R_{\pi }(0){\times }R_{\nu }(0)$ &               
$R_{\pi }(\pi ){\times }R_{\nu }(0)$ &               
  $R_{\pi }(0){\times }R_{\nu }(\pi )$ &
 $R_{\pi }(\pi ){\times }R_{\nu }(\pi )$  \\
\cline{1-4}
$H_0$ & $H_0-10(k{+}2k_6)N$ & $H_0-10(k{+}2k_6)N$ 
& $H_0$ \\
\cline{1-4}
$k_1^{\pi}$ & 
$k_1^{\pi}{+}2k_6^{\pi}{+}4(N_{\pi}{+}2)(2k_6{+}k)\!\!$ & 
$k_1^{\pi}{+}4(N_{\pi}{+}2)(2k_6{+}k)$  &
$k_1^{\pi}{+}2k_6^{\pi}$  \\
$k_2^{\pi}$ &
$k_2^{\pi}{+}2k_6^{\pi}{-}8k_6{-}4k$ &
$k_2^{\pi}{-}8k_6{-}4k$  &
$k_2^{\pi}{+}2k_6^{\pi}$  \\         
$k_3^{\pi}$ &
$k_3^{\pi}{-}6k_6^{\pi}{-}2k_5{+}2k_6{-}8k_7$ &
$k_3^{\pi}{-}2k_5{+}2k_6{-}8k_9$  &
$k_3^{\pi}{-}6k_6^{\pi}{-}8k_8$ \\        
$k_4^{\pi}$ &
$k_4^{\pi}{+}2k_6^{\pi}$ &
$k_4^{\pi}$ & 
$k_4^{\pi}{+}2k_6^{\pi}$ \\
$k_5^{\pi}$ &
$k_5^{\pi}{+}4k_6^{\pi}{+}2k_5{-}2k_6{+}8k_7$  &
$k_5^{\pi}{+}2k_5{-}2k_6{+}8k_9$ &
$k_5^{\pi}{+}4k_6^{\pi}{+}8k_8$ \\
$k_6^{\pi}$ &
${-}k_6^{\pi}$ &
$k_6^{\pi}$ &
${-}k_6^{\pi}$ \\
\cline{1-4}
$k_1^{\nu}$ &
$k_1^{\nu}{+}4(N_{\nu}{+}2)(k{+}2k_6)$ &
$k_1^{\nu}{+}2k_6^{\nu}{+}4(N_{\nu}{+}2)(k{+}2k_6)\!\!$ &
$k_1^{\nu}{+}2k_6^{\nu}$ \\
$k_2^{\nu}$ &
$k_2^{\nu}{-}8k_6{-}4k$ &
$k_2^{\nu}{+}2k_6^{\nu}{-}8k_6{-}4k$ &
$k_2^{\nu}{+}2k_6^{\nu}$ \\
$k_3^{\nu}$ &
$k_3^{\nu}{-}2k_5{+}2k_6{-}8k_7$ &
$k_3^{\nu}{-}6k_6^{\nu}{-}2k_5{+}2k_6{-}8k_9$ &
$k_3^{\nu}{-}6k_6^{\nu}{-}8k_8$ \\
$k_4^{\nu}$ &
$k_4^{\nu}$ &
$k_4^{\nu}{+}2k_6^{\nu}$ &
$k_4^{\nu}{+}2k_6^{\nu}$ \\
$k_5^{\nu}$ &
$k_5^{\nu}{+}2k_5{-}2k_6{+}8k_7$ &
$k_5^{\nu}{+}4k_6^{\nu}{+}2k_5{-}2k_6{+}8k_9$ &
$k_5^{\nu}{+}4k_6^{\nu}{+}8k_8$ \\        
$k_6^{\nu}$ &
$k_6^{\nu}$ &
${-}k_6^{\nu}$ &
${-}k_6^{\nu}$ \\
\cline{1-4}
$k$ & $k$ & $k$ & $k$  \\         
$k_2$ & $k_2$ & $k_2$ & $k_2$   \\       
$k_3$ & $k_3{+}2k_5{+}2k_6{+}2k{+}8k_7$ &
$k_3{+}2k_5{+}2k_6{+}2k{+}8k_9$ & $k_3{+}8k_8$ \\
$k_4$ & $k_4$ & $k_4$ & $k_4$  \\         
$k_5$ & ${-}k_5{-}k{-}8k_7$ & ${-}k_5{-}k{-}8k_9$ & 
$k_5{-}8k_8$  \\  
$k_6$  & ${-}k_6{-}k$ & 
${-}k_6{-}k$ & $k_6$  \\
$k_7$ & $k_7$ & $k_8{+}k_9$ & 
$k_8{+}k_9$ \\         
$k_8$  & ${-}k_7{+}k_9$ & 
$k_7{-}k_9$ & ${-}k_8$ \\    
$k_9$  & $k_7{+}k_8$ &
$k_9$ & $k_7 {+}k_8$ \\
\cline{1-4}
\end{tabular}
\end{center}
\end{table}
In the case (iii) we do not immediately obtain parameter symmetries.
However, for any possible transformation $R_\rho(\varphi_\rho)$ [or 
$\tilde{R}_\rho(\varphi_\rho)$] there always can be found some constraints on 
the parameters $\{k^\pi_i,k^\nu_i,k_i\}$ such that the 
Hamiltonians $H(\{k^\pi_i,k^\nu_i,k_i\})$ and 
$R_\rho(\varphi_\rho)\,H(\{k^\pi_i,k^\nu_i,k_i\})$ [or 
$\tilde{R}_\rho(\varphi_\rho)\,H(\{k^\pi_i,k^\nu_i,k_i\})$]
become isospectral 
even if there is no isospectrality between these Hamiltonians in the
general case. Hence in the case (iii) we obtain additional
non-standard parameter symmetries that are valid only if the parameters fit 
some relations. These additional symmetries are listed in Table~3 (the
constraining relations for the parameters are given in the first row).

All IBM-2 parameter symmetry relations with the only exception of the 
parameter symmetry D (see Table~2), involve the total number of proton 
bosons $N_\pi$ or/and the total number of neutron bosons $N_\nu$ 
($N=N_\pi+N_\nu$). Hence there is a principal possibility to distinguish
between few parameter sets giving rise to identical spectra by the 
analysis of the spectra of neighboring isotopes or/and isotones.  

\begin{table}
\epsfig{file=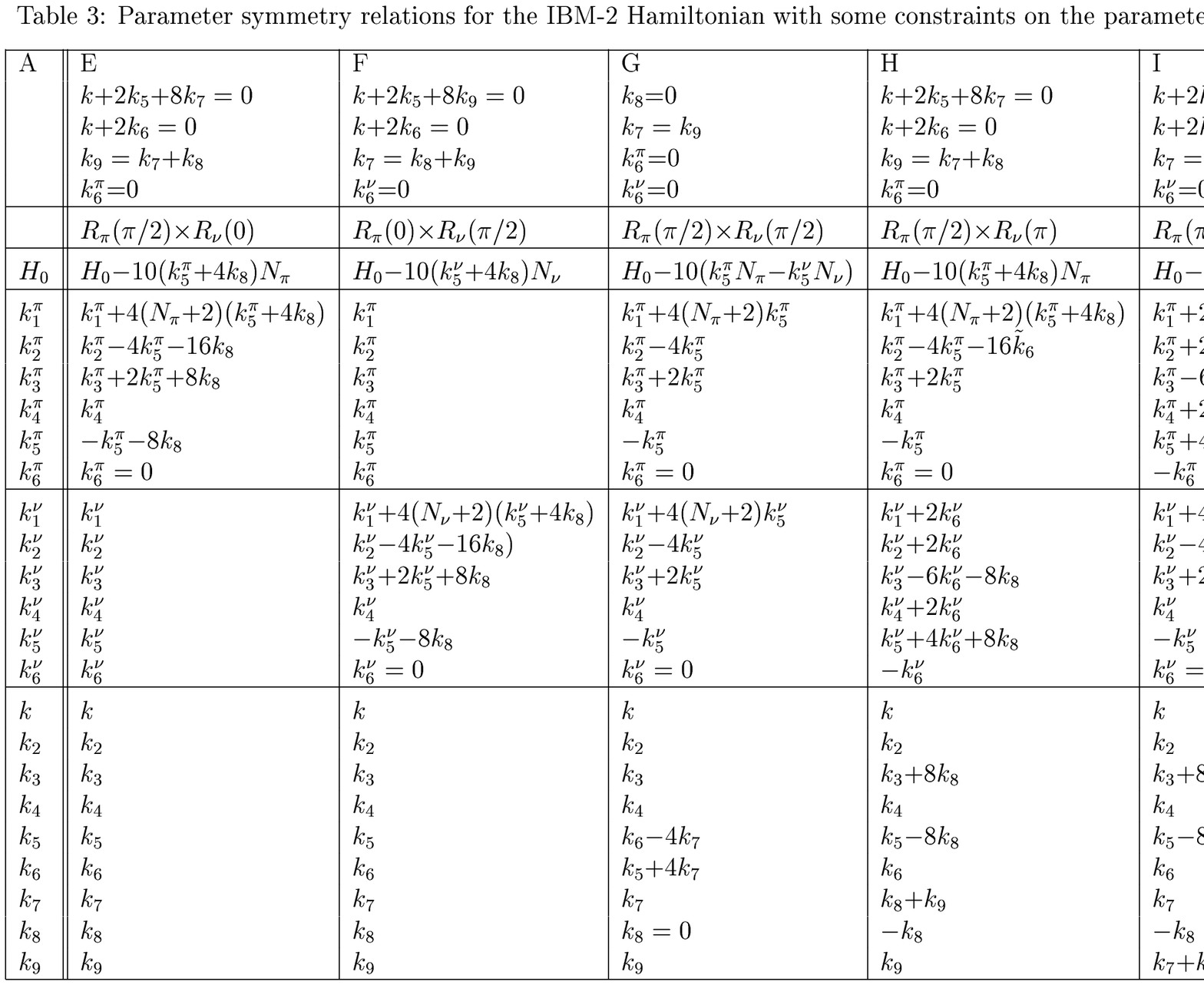,angle=90,height=0.99\textheight}
\end{table}

%


\section{Summary}
We have analyzed canonical transformations of boson operators consistent
with the structure of boson Hamiltonian in the cases of IBM-1 and IBM-2,
or, equivalently, different realizations of the symmetry algebra
of the model and its subalgebras. Analysis of alternatively embedded
subalgebras is of a particular importance  in the case of IBM-2,
because it provides a regular way to construct the general IBM-2 
Hamiltonian.

One can suppose that there should be no signals in physical applications 
of switching from one realization of the symmetry 
algebra to another equivalent realization. 
However, it is not so. The existence of alternative realizations of the 
symmetry algebra manifests itself as parameter symmetries of the model, 
i.e. as existence of few sets of Hamiltonian parameters providing 
identical spectra. The parameter symmetry is of physical importance 
for applications of IBM since the parameters of the model are obtained 
by the fit to experimental spectra. We have shown that in some cases one
can discriminate between the sets of parameters related by the parameter 
symmetry by analyzing spectra of neighboring isotopes and/or isotones. 

\par
The parameter symmetry is a common property of boson models.
For example, we have shown~\cite{VM} that it is present in the vibron 
model (see e.g.~\cite{Frank}); we suppose that it can be found in 
$sdg$-IBM and other algebraic models including fermion and boson-fermion 
ones. There can exist other possibilities of discriminating between 
isospectral parameter sets (see, 
e.g., Ref.~\cite{Yoshida} where the proton-neutron interacting 
boson-fermion model is discussed). 

Before finishing the paper, we mention few recent papers related to the 
present investigation. D.~Kusnezov~\cite{Kusnezov} discussed in detail
hidden symmetries, i.e. the particular cases of parameter symmetries
relating transitional Hamiltonians to the ones corresponding to  DS
limits. He noted the relevance of hidden symmetries to the studies of
chaos. A more detailed discussion of this item can be found in
Ref.~\cite{Jolie}. In Ref.~\cite{d-parity} the so-called $d$ parity was
introduced for IBM Hamiltonian in the U(5)--SO(6) transitional case. The
$d$ parity operator commutes  with the Hamiltonian and provides  an
additional quantum number for qualification of the energy levels,
electromagnetic transitions, etc.
\par
We are thankful to  R.~Bijker, D.~Bonatsos, C.~Daskaloyannis,
G.~F.~Filippov, F.~Iachello, R.~V.~Jolos, V.~P.~Karassiov, T.~Otsuka,
J.~Patera, N.~Pietralla, D.~L.~Pursey, V.~N.~Tolstoy and P.~Van~Isacker
for valuable discussions. The paper is supported partly by
the Competitive Center at St.~Petersburg State University,
the Russian Foundation of Basic Research, and
the European Community through project CI1*-CT94-0072.


\end{document}